# Giant spin-phonon-electronic coupling in a 5d oxide

S. Calder,[1,*] J. H. Lee,[2,*] M. B. Stone,[1] M. D. Lumsden,[1] J. C. Lang,[3] M. Feygenson,[4], Y. G. Shi,[5, 6] Y. S. Sun,[6,#] Y. Tsujimoto,[7] K. Yamaura,[6, 8] and A. D. Christianson[1,9]

[1]Quantum Condensed Matter Division, Oak Ridge National Laboratory, Oak Ridge, Tennessee 37831, USA.
[2]Material Science and Technology Division, Oak Ridge National Laboratory, Oak Ridge, Tennessee 37831, USA.
[3]Advanced Photon Source, Argonne National Laboratory, Argonne, Illinois 60439, USA.
[4]Chemical and Engineering Materials Division, Oak Ridge National Laboratory, Oak Ridge, Tennessee 37831, USA.
[5]Institute of Physics, Chinese Academy of Sciences, Beijing 100190, China.
[6]Superconducting Properties Unit, National Institute for Materials Science, 1-1 Namiki, Tsukuba, Ibaraki 305-0044, Japan.
[7]Materials Processing Unit, National Institute for Materials Science, 1-2-1 Sengen, Tsukuba, Ibaraki 305-0047, Japan.
[8]Graduate School of Chemical Sciences and Engineering, Hokkaido University, North 10 West 8, Kita-ku, Sapporo, Hokkaido 060-0810, Japan.
[9]Department of Physics and Astronomy, University of Tennessee, Knoxville, Tennessee 37996-1200, USA.
[#] Current address (Ying Sun): Center of Condensed Matter and Materials Physics, Beihang University, Beijing 100191, China.

[*]e-mail: caldersa@ornl.gov; leej@ornl.gov

**Enhanced coupling of material properties offers new fundamental insights and routes to multifunctional devices. In this context 5d oxides provide new paradigms of cooperative interactions driving novel emergent behavior. This is exemplified in 5d osmates that host a metal-insulator transition (MIT) driven by magnetic order. Here we consider the most robust case, the 5d perovskite $NaOsO_3$, and reveal a giant coupling between spin and phonon through a frequency shift of $\Delta\omega=40$ cm$^{-1}$, the largest measured in any material. We identify the dominant octahedral breathing mode and show isosymmetry with spin ordering which induces dynamic charge disproportionation that sheds new light on the MIT. The occurrence of the dramatic spin-phonon-electronic coupling in $NaOsO_3$ is due to a property common to all 5d materials: the large spatial extent of the 5d ion. This allows magnetism to couple to phonons on an unprecedented scale and consequently offers multiple new routes to enhanced coupled phenomena.**

The coupling, tuning and enhancement of distinct and finely balanced collective phenomena in materials via an external perturbation provide ideal mechanisms for innovative multifunctional applications and constitutes a central theme of material science research. Consequently significant scientific effort is devoted to uncovering new classes of materials with novel and coupled properties. In this context 5d transition metal oxides appear promising candidates since they host a unique balance of electronic properties of extended wavefunctions, strong spin-orbit coupling (SOC) and large crystalline electric field (CEF) splitting that are all of comparable energy (~1-4 eV) in which cooperation can, and often does, lead to novel behavior emerge not observed elsewhere in the periodic table [1,2]. This is highlighted in the distinct and exotic insulating states of iridates and osmates. For example in $Sr_2IrO_4$,[3] and other iridates, [1,2] SOC alters the electronic ground state to create a condition that allows even the small on-site Coulomb interaction to force an insulating band gap. The result is a new route to a Mott-like insulator. Conversely, the neighboring osmate $NaOsO_3$ that we focus on here hosts the most robust example of a Slater metal-insulator transition (MIT) that falls outside the Mott-Hubbard paradigm [6-9]. In the case of a Slater MIT the periodic potential created by the magnetic order acts as a direct and continuous tuning parameter between metallic and insulating states.

In 5d oxides the electronic properties, particularly SOC and extended electronic wavefunctions, are expected to lead to a much greater coupling of magnetic and electronic phenomena to the crystalline lattice compared to the much-studied 3d transition metal analogues. This would thereby provide new pathways for enhanced material functionality and offer the potential for new fundamental insights, yet this remains a largely unexplored assertion that we address, and support, in this investigation. We probe phonon modes that are a manifestation of collective excitations in a crystalline lattice and offer insights into material properties and associated couplings. For example spin-phonon coupling has been a fertile ground in multiferroic investigations [10,11] of materials with the same perovskite structure as $NaOsO_3$, as well as in a variety of materials including high temperature superconductors.[12] The largest reported phonon shift in a perovskite is found in $(Sr,Ba)MnO_3$ with a value of $\Delta\omega=25 cm^{-1}$ in the $TO_1$ polar phonon.[13,14] In the context of 5d materials recently it has been shown that the mixed 3d-5d half-metal double perovskite $Ba_2FeReO_6$ hosts a dramatic spin-electron-phonon coupling as evidenced by a phonon shift of $\Delta\omega=30 cm^{-1}$,[15] the largest ever reported prior to our present work on $NaOsO_3$. The phonon shift in $Ba_2FeReO_6$ occurs simultaneously with a structural symmetry change thereby complicating the interpretation, however the behavior is reported as being directly linked to the interaction between the 3d and 5d ions. Conversely, we show here that 5d ions alone produce even larger spin-phonon shifts and are the key ingredient to attain large couplings between magnetic, structural and electronic degrees of freedom. In NaOsO$_3$ the giant phonon shift occurs without a structural symmetry change [7,8] and any subtle local symmetry changes through the Slater MIT have been ruled out here with detailed neutron pair density functional (PDF) measurements, see Supplementary Fig S1. While a symmetry change does not occur, the lattice reflects the magnetic MIT with a small shift in the a and c lattice constants of under 0.1%, as previously reported.[8] This change would appear to be of only minimal consequence in most materials. As we discuss, however, this reveals an unanticipated and critical sign of strong coupling in 5d materials due to the inherent extended orbitals that can act as signature to extend investigations to further 5d systems.

**Results**

**Giant spin-phonon coupling measured with inelastic neutron scattering.** To follow the behavior of collective excitations in NaOsO$_3$ through the magnetic MIT we performed inelastic neutron scattering (INS) measurements. Figure 1**a** shows the key result of the temperature dependence of the phonon density of states (pDOS) whose peaks are related to the underlying phonon modes for the region around 700 cm$^{-1}$ that covers the essential physics of interest. The full spectrum is shown in Supplemental Fig S2. Three distinct peaks in the pDOS are observed around 700 cm$^{-1}$ and fitting these each to a Gaussian, as shown, allows the energy of the modes to be followed with temperature. A dramatic phonon shift is observed as revealed by the frequency shift of the centre of the modes in Fig 1**a-b**. Moreover there is an anomalous and counterintuitive intensity increase with decreasing temperature through the MIT as shown inset Fig 1**a** considering the entire range of 550 to 800 cm$^{-1}$. The results are significant in several regards and go beyond observations in any previous system. Firstly, the onset of the anomalous phonon mode shift is concurrent with the reported magnetic MIT in NaOsO$_3$ at 410 K, indicating a coupling of the phonons to the magnetic and electronic transitions. Secondly, the phonons show a remarkable "giant" shift of $\Delta\omega=40$ cm$^{-1}$, the largest measured in any material for a spin-phonon coupled transition.

**Theoretical demonstration of giant spin-phonon shift.** To gain a fundamental insight into the origin of the anomalous phonon mode behavior in NaOsO$_3$ and disentangle the myriad of competing interactions at the magnetic and electronic transition we performed detailed density functional theory (DFT) calculations. The calculations predict the same three phonon modes observed with INS between 600 to 900 cm$^{-1}$, see Fig 1**c**, and as expected for the orthorhombic structure in NaOsO$_3$ these are themselves composed of three branches, unresolvable in the current powder INS measurement. The calculated shift is in agreement with the record measured value of $\Delta\omega=40$cm$^{-1}$. Our calculations obtain all of the corresponding phonon modes allowing us to access the underlying physics driving the behavior. The modes are shown in Fig 1**d**, and correspond to Os-O vibrations, specifically they are breathing modes B$_{1g}$ (in-phase) and B$_{2g}$ (out of phase) and two asymmetric Jahn-Teller stretching modes A$_g$ (in phase) and B$_{3g}$ (out of phase).

**Extended 5d electronic wavefunction responsible for enhanced coupling.** All of the four modes uncovered (B$_{1g}$, B$_{2g}$, A$_g$ and B$_{3g}$) are characterized by simultaneous bond-stretching/shrinking between the oxygen and osmium ions of the OsO$_6$ octahedra.[16] The direct and mutual consequence of a change in any of the phonon modes is therefore an alteration of the Os-O wavefunction overlap, with this alteration directly changing the magnetic superexchange that is mediated by the oxygen ions. Since 5d oxides have an intrinsically large wavefunction overlap, the deviation of the overlap of the oxygen displacements is magnified to a much greater extent than in other systems, such as the much-studied 3d transition metal oxides. As we observe, the result is that as the magnetic order, that develops and drives the Slater MIT, can couple to the lattice on an unprecedented scale and here manifests in the observed giant spin-phonon coupling in NaOsO$_3$. Moreover this suggests that in general an exploration and reexamining of 5d materials that host subtle lattice anomalies at magnetic phase transitions should reveal similar enhanced coupled phenomena. While SOC is often attributed to anomalous behavior of 5d materials, in NaOsO$_3$ the 5d$^3$ t$_{2g}$-degenerate ground state will suppress the effective SOC.[17] We nevertheless addressed the role of SOC with x-ray absorption near edge spectroscopy (XANES) that allows a quantitative comparison with SOC enhanced iridates. As expected our results indicate SOC does not play a dominant role in the behavior of NaOsO$_3$ as discussed in the supplementary material (Note 1 and Fig. S2). Our first-principle results additionally show the large coupling without SOC. Therefore the extended electronic 5d wavefunction that magnifies and couples to small lattice distortions is the key ingredient to the giant spin-phonon coupling observed. Indeed since spin-phonon coupling is an emergent phenomena required to design functional materials,[18] the potential for inherently large spin-phonon coupling in 5d oxides can provide novel routes for multifunctional devices. Specifically our results indicate 5d materials with subtle lattice anomalies concurrent with a separate phase transition should be reexamined in a new light.

**Octahedral B$_{2g}$ breathing mode is the dominant mode.** While we have been able to explain the anomalous spin-phonon behavior in the general context of 5d materials we are also able to interpret the role of the spin-phonon behavior in NaOsO$_3$ and reveal the underlying mechanisms that control and drive the exotic properties that emerge at the Slater transition by considering the separate phonon modes. It is instructive to first consider the static behavior of the octahedra and propensity towards Jahn-Teller distortion. This can be quantified by parameters $Q_2$ and $Q_3$, which are shown schematically in Fig 2**a**, and defined as $Q_2 = (X_1-X_4-Y_2+Y_5)/\sqrt{2}$ and $Q_3=(2Z_3-2Z_6-X_1+X_4-Y_2+Y_5)/\sqrt{6}$, where X,Y,Z are the oxygen positions. [19] Thereby the values of $Q_2$ and $Q_3$ reveals the degree of static octahedral anisotropy, with the larger the value the more distorted the octahedra. Calculations from experimentally determined atomic parameters for NaOsO$_3$ [3] reveal $Q_2$ and $Q_3$ to be small at all temperatures, but counterintuitively decrease through the Slater MIT. Specifically, at 500 K $Q_2$=0.0114(15) a.u. and at 300 K $Q_2$=0.0035(11) a.u.. While at 500 K $Q_3$=0.0171(18) a.u and at 300 K $Q_3$=0.0114(14) a.u. Therefore this reveals that in NaOsO$_3$ the octahedra become more isotropic, both in the *xy*-plane and in three dimensions (3D), at low temperature in the insulating regime. This behavior is at odds to the normal Jahn-Teller distortions of increased anisotropy and does not favour the asymmetric stretching modes $A_g$ and $B_{3g}$. Instead the increased static isotropy is more conducive to the symmetric breathing distortions $B_{1g}$ and $B_{2g}$. Indeed the abnormal behavior of the intensity increase of the pDOS, inset Fig. 1**a**, is consistent with an increase in vibration with decreasing temperature, counter to usual thermal behavior. This appears most pronounced at the highest frequency, which corresponds to the breathing mode $B_{2g}$. The increased octahedral isotropy will therefore favour and promote this 3D breathing distortion thereby leading to an increase in amplitude of the distortion below the MIT resulting in the abnormal intensity increase. Collectively the anomalous intensity behavior and large frequency shift, that appears most pronounced at the highest frequency, allows us to assign the $B_{2g}$ breathing distortion as central to the behavior in NaOsO$_3$.

**Isosymmetric ordering of magnetic structure and breathing mode cooperate within the insulating regime.** The anomalous phonon behavior and the assignment of the dominant $B_{2g}$ breathing mode now provides a complete picture of the emergence of the giant spin-phonon-electronic coupling in NaOsO$_3$ at the Slater MIT. Unexpectedly the $B_{2g}$ mode can actually open the insulating gap, even in the paramagnetic regime, see Fig. 2**b**. Our frozen DFT results show that the gap opening occurs directly due to the 3D octahedral breathing ordering in the perovskite structure as neighboring octahedra expand/contract and create a periodic charge disproportionation on the Os ion. We note that no other phonon distortions produce similar ordering or open a gap. The minimum required oxygen displacement is u=0.2 Å of the $B_{2g}$ mode to create a gap, see Fig. 2**b**. This corresponds to 10% of the actual Os-O bond distance and therefore is too large to allow this mechanism alone to drive the MIT. As shown in Fig 2**c**, that substantiates earlier work, [7] it is the onset of antiferromagnetic order alone that creates the insulating gap via the Slater mechanism in $NaOsO_3$. However, as shown schematically in Fig. 3, the G-type AFM and $B_{2g}$ mode are isosymmetric: the G-type magnetic order has every nearest neighboring spin oppositely aligned while the $B_{2g}$ mode has every nearest neighboring octahedra oppositely expanded/compressed. This manifests in cooperative behavior in the insulating regime, indeed both create the same zone folding that can result in a gap formation in $NaOsO_3$. Statically, even in the paramagnetic regime, the octahedral breathing creates a strong charge disproportionation of Dd/Du=7.0e/Å in the lattice due to the maximally (3D) change of the electronic potential around the Os ion and places the system on the verge of a MIT. Considering a nominal phonon vibration displacements of the order 0.01Å, the dynamic charge disproportionation will be ~0.01e. The consequence of the isosymmetric spin/octahedral ordering is a cooperation between magnetic order and the lattice that drives increased octahedral isotropy to accommodate and promote the breathing distortion and increase the insulating gap. The result is the observed anomalous phonon frequency and intensity behavior we revealed in Fig1**a**. Indeed this indicates that control of the octahedra via pressure or strain is a route to tune the MIT in $NaOsO_3$. Such a handle would apply even in the paramagnetic regime and indicates the potential for further cooperative phenomena.

**Coupling of lattice, magnetic order and MIT.** The remarkable coupled properties in $NaOsO_3$ are illustrated in Fig. 4 where we experimentally show that there is a direct scaling of the structural anomaly of the lattice constants, phonon mode shift, magnetic moment, with the MIT qualitatively following a similar trend. While the realization of numerous overlapping phenomena is currently rare it is likely that additional 5d materials will host similar rich phase diagrams. Specifically investigating 5d materials with small lattice alterations at magnetic phase transitions are a fruitful arena for exploration of coupled properties, with the extended 5d wavefunctions the key factor in allowing the large coupling of the lattice to the magnetic or electronic material properties.

# Discussion

The measurement and interpretation of a "giant" spin-phonon mode shift of $\Delta\omega=40$ cm$^{-1}$ in the 5d perovskite NaOsO$_3$, the largest observed in any material, reveals materials with 5d transition metal ions as providing an exceptional route to achieve large coupling between different material properties, such as magnetism, lattice and electrical resistance. The enhanced coupling is directly related to the extended electronic wavefunctions of 5d ions that allows extremely small changes of the lattice, 0.1% in NaOsO$_3$, to couple to magnetic or electronic degrees of freedom on a giant scale. Consequently the reported results are applicable in general to 5d materials, which is of particular interest due to the presence of new paradigms of finely balanced competing interactions in 5d systems that result in novel emergent magnetic and insulating states. For example 5d systems offer new routes to a Mott insulator in iridates due to strong SOC, Weyl semi-metal behavior, potential avenues to high temperature superconductivity and topological insulators. The consequence of associated enhanced coupling is therefore intriguing. In the case of NaOsO$_3$ we find the magnetic order, that drives a magnetic Slater metal-insulator transition, couples in a dramatic way to the phonons resulting in the observed record high phonon shift and spin-phonon-electronic coupling. Not only is the behavior we reveal on an unprecedented scale, but gaining an understanding of the coupling has provided new insights into the underlying exotic physics at the phase transition by revealing that phonon-driven isotropy and strong charge disproportionation cooperate within the magnetic Slater mechanism to open the insulating gap.

Beyond the specific behavior of NaOsO$_3$ the central result of the giant coupling is to sit in the wider context of 5d compounds by revealing an unprecedented example of the potential for enhanced coupling from remarkably small lattice alterations that would not occur in a similar 3d transition metal oxide due to the relatively limited spatial extent of the 3d wavefunction and consequently offers new directions to multifunctional devices as well as fundamental insights. Further research into 5d materials is therefore a promising avenue to achieve large enhanced coupling where small lattice anomalies are observed and also offers routes to coupled phenomena of properties not possible elsewhere in the periodic table due to the unique properties of 5d ions.

**Methods**

**Synthesis.** Polycrystalline samples of NaOsO$_3$ were prepared using a high pressure solid state synthesis with pressures of 6 GPa, as described in Ref. 7.

**Inelastic neutron scattering.** Inelastic neutron scattering measurements on the ARCS spectrometer at the Spallation Neutron Source (SNS) were performed on a 5g polycrystalline sample of NaOsO$_3$. The sample was loaded into a vanadium can and measurements performed between 300 K to 500 K using an incident energy of $E_i$ = 120 meV. Corrections for the Bose factor, where appropriate, were performed using the DAVE software.

**DFT.** First-principles calculations were performed using density-functional theory within the generalized gradient approximation GGA+U method [6] with the Perdew-Becke-Erzenhof parameterization as implemented in the Vienna ab-initio Simulation Package (VASP 5.3). [19] Theoretical details for spin-phonon coupling are described in Ref. 16. We use the Dudarev [20] implementation with on-site Coulomb interaction U=1.7eV and on-site exchange interaction $J_H$=1eV, so $U_{eff}$=0.7 eV to treat the localized d electron states in Os. Within GGA+U, this small U gives excellent agreement between the experimental Neel temperature ($T_N$ =411K) and calculated one ($T_{N,MFT}$ =415 K) in mean-field approximation. The projector augmented wave (PAW) potentials [22] explicitly include 9 valence electrons for Na ($2s^2\ 2p^6\ 3s^1$ ), 14 for Os ($5p^6\ 5d^5\ 6s^2$), and 6 for oxygen ($2s^2 2p^4$). To capture spin-phonon coupling w.r.t. temperature, we employed the method successfully used for various magnetic perovskites.

**Neutron pair density function**. nPDF measurements were performed on the NOMAD beamline at the SNS on a powder sample from 370 K to 460 K. The data was analyzed and modeled with pdfgui.

**XANES.** X-ray absorption measurements were performed at the Advanced Photon Source on sector 4-ID-D. Spectra were collected at room temperature on a powder sample (~100mg) in transmission mode through the Os L$_2$ and L$_3$ edges. Analysis was performed with the Athena software.

**Acknowledgements**

The research at ORNL's Spallation Neutron Source was supported by the scientific User Facilities Division, Office of Basic Energy Sciences, U.S. Department of Energy (DOE) was sponsored by the Scientific User Facilities Division, Office of Basic Energy Sciences, U.S. Department of Energy. Use of the Advanced Photon Source, an Office of Science User Facility operated for the U.S. DOE Office of Science by Argonne National Laboratory, was supported by the U.S. DOE under Contract No. DE-AC02-06CH11357. Research was supported in part by Grant-in-Aid for Scientific Research (22246083, 22850019, 25289233) from JSPS and FIRST Program from JSPS and ALCA program from JST and the Ministry of Science and Technology of China (973 Project No. 2011CBA00110). This research was supported by National Natural Science Foundation of China (No. 11274367)).

**Figure legends**

**Figure 1 | Measured and calculated phonon modes in NaOsO$_3$ reveal a giant shift at the magnetic MIT. (a)** Thermal evolution of the phonon mode density of states measured with inelastic neutron scattering through the Slater MIT temperature of 410 K. Three modes are resolvable between 550 and 800 cm$^{-1}$ at all temperatures. These were each modeled to a Gaussian lineshape with the width of the energy resolution (14 cm$^{-1}$ FWHM). The different temperature measurements are offset in intensity to aid comparison. The three vertical lines indicate the centre of the Gaussian fits to the spectra for each temperature. Inset reveals an abnormal intensity increase with decreasing temperature of the integrated intensity over the region 550 to 800 cm$^{-1}$. **(b)** The measured phonon mode frequency centres obtained from inelastic neutron scattering and **(c)** from DFT calculations both show strong agreement and reveal a giant phonon shift at the Slater transition of Δω=40 cm$^{-1}$. The DFT calculations allow assignment of the responsible modes, as indicated. The breathing modes occur at higher frequencies than the asymmetric stretching, with B$_{2g}$ occurring at the highest frequency. **(d)** The separate distortions are shown. A$_g$ (in phase) and B$_{3g}$ (out of phase) correspond to asymmetric stretching. B$_{1g}$ (in phase) and B$_{2g}$ (out of phase) represent symmetric stretching breathing modes.

**Figure 2 | The dominant B$_{2g}$ breathing mode provides a route to an insulating band gap in NaOsO$_3$. (a)** The static octahedral distortion can be quantified with parameters Q$_2$ and Q$_3$ that represent the degree of octahedral anisotropy. Unusually Q$_2$ and Q$_3$ both show reduced values below the Slater MIT, compared to high temperature. At 500 K Q$_2$=0.0114(15) a.u. while at 300 K Q$_2$=0.0035(11) a.u.. Similarly At 500 K Q$_3$=0.0171(18) a.u. while at 300 K Q$_3$=0.0114(14) a.u.. This indicates the octahedra actually become more isotropic at lower temperature. **(b)** The increased octahedral isotropy is a consequence of the B$_{2g}$ distortion increasing in the low temperature insulating regime, leading to the anomalous measured pDOS intensity increase. Intriguingly this three dimensional breathing distortion can open a band gap. This is revealed in the DOS from DFT calculations when a displacement of u=0.2 Å is frozen into the structure, where u denotes the displacement of each oxygen. **(c)** The required breathing mode displacement is too large to solely account for the observed MIT in NaOsO$_3$, instead the G-type magnetic ordering alone can open the gap via the Slater mechanism.

**Figure 3 | G-type magnetic order and breathing mode couple within the insulating regime.** At the Slater MIT NaOsO$_3$ forms G-type antiferromagnetic ordering. This magnetic order (blue arrows), with every nearest neighboring spin oppositely aligned in three dimensions, follows the same ordering as the B$_{2g}$ breathing mode expansion/contraction and subsequent charge disproportionation, δ, (magenta sphere) on the Os ion revealed from DFT. The isosymmetric magnetic order and breathing distortion ordering both cooperate within the insulating regime. The result is the observed giant spin-phonon-electronic coupling in NaOsO$_3$. The value of charge disproportionation, δ(e), is indicated in the schematic and calculated for frozen breathing mode distortions, u(Å).

**Figure 4 | Phonon, lattice and magnetic degrees of freedom couple through the Slater MIT in NaOsO$_3$.** The measured temperature dependence of the phonon mode frequency shift, the (110) magnetic Bragg reflection intensity (Ref. 3) and the *a-c* lattice constants in NaOsO$_3$ (altered from Ref. 3) show a direct scaling with temperature through the Slater MIT due to spin-phonon-electronic coupling. This coupling and enhancement of several material properties indicates new routes to novel functional materials with 5d ions, where the extended electronic wavefunction can magnify phenomena from small lattice anomalies of ~0.1%. The lattice parameters have been corrected by removing a constant sloping thermal background.

**Figure 1**

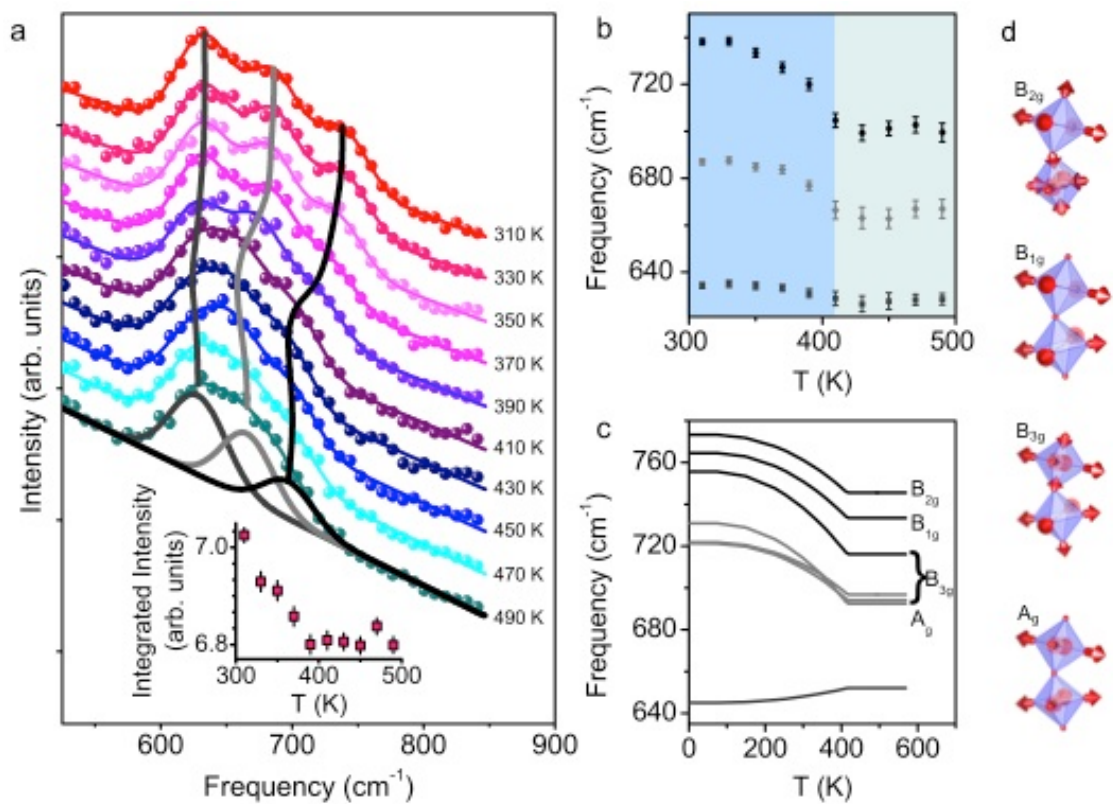

**Figure 2**

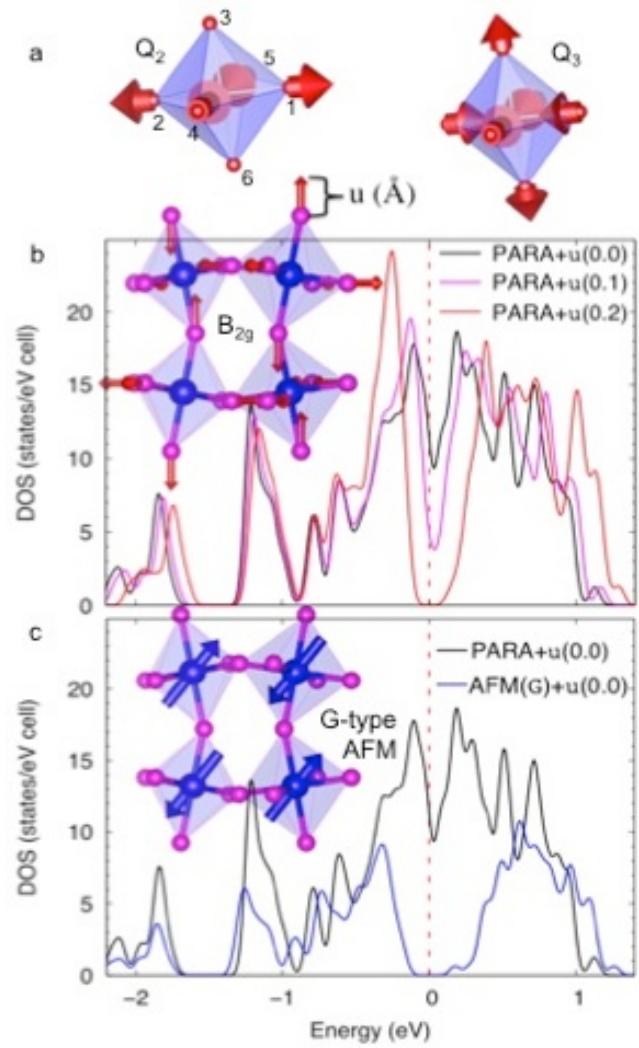

**Figure 3**

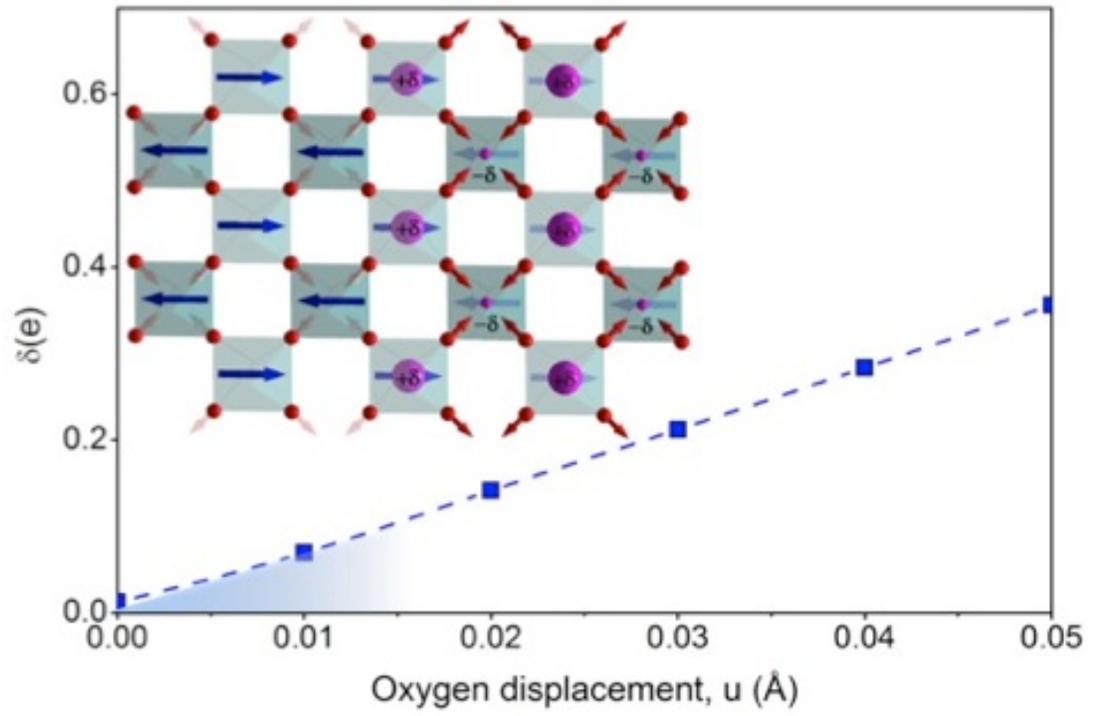

**Figure 4**

Supplemental Material

Supplemental Figure S1

**S1 | Neutron PDF measurements through the magnetic MIT transition indicate no local symmetry change.** Neutron PDF results from 370 K to 464 K data are offset in intensity and fit to the Pnma structure using the PDFgui software. The results reveal no local symmetry change occurs through the 410 K Slater transition in $NaOsO_3$. Therefore on both a local and global level $NaOsO_3$ undergoes no structural symmetry in the temperature range of interest.

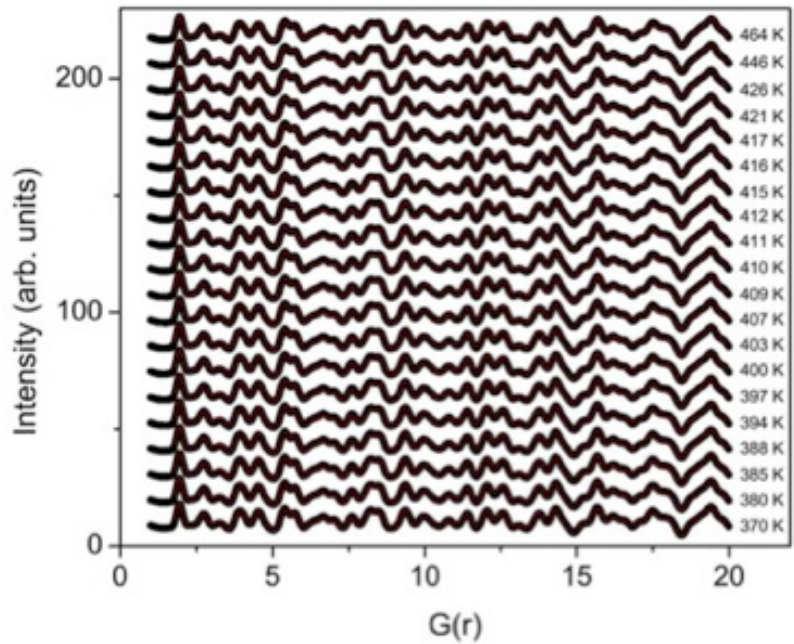

**Supplemental Figure S2**

**S2 | Full inelastic neutron scattering spectra above and below the transition.** Inelastic neutron scattering measurements with an $E_i$=120 meV over the full range of energy through the temperature region of interest. The only energy at which the intensity increases at low temperature is around the frequency we have focused on, ~750 cm$^{-1}$. The data is plotted without correcting for the Bose factor. The large intensity variation at low frequency, particularly around 100 cm$^{-1}$, is expected and is a consequence of increased thermal population with increasing temperature.

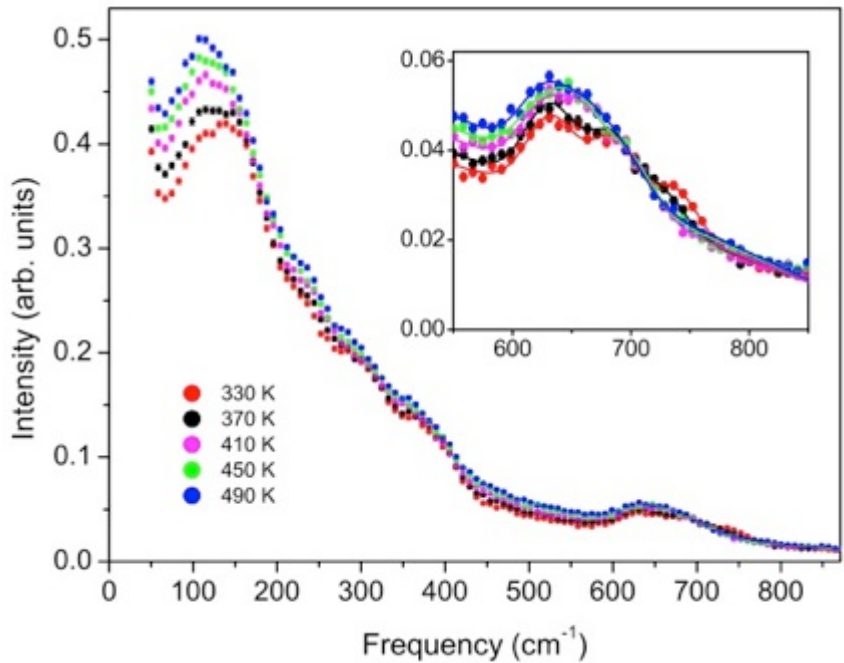

**Supplemental Note 1**

**Role of SOC in NaOsO$_3$.** The significant enhancement of SOC in going from 3d to 5d ions offers a potential explanation for anomalous behavior between 3d and 5d analogues. Previous experimental and theoretical investigations, however, have found that SOC does not play a dominant role in the electronic ground state of NaOsO$_3$,[4,5] distinct from the neighboring iridates. Nevertheless SOC is often employed and cited as a significant factor in general in 5d oxides and therefore we directly addressed the role on a quantitative level to rule out SOC as being a principle factor at play in the anomalous phonon behavior in NaOsO$_3$. An important experimental tool in this regard to investigate the role of SOC, and allow comparisons between different materials and ions, is x-ray absorption near edge spectroscopy (XANES). Results of such measurements on NaOsO$_3$ are shown in Fig. S2. A statistical L3:L2 white-line branching ratio BR of 2:1 is expected in XANES, independent of the electron occupancy of the ion under investigation. The measured ratio of integrated intensities for the L3:L2 edge is 2.6:1. This allows a direct comparison of Os$^{5+}$ in NaOsO$_3$ with Ir$^{4+}$ in Sr$_2$IrO$_4$ and BaIrO$_3$ that show significant SOC effects and additionally with other 5d ions, such as Re, that show reduced SOC effects. The BR of 2.6:1 in NaOsO$_3$ contrasts with 4:1 for BaIrO$_3$, with the large deviation from 2:1 presented as direct proof of strong spin-orbit coupling.[6] To directly compare our results we extract the ground state expectation value of the angular part of the spin-orbit coupling $<\mathbf{L.S}>$, through BR = (2+r)/(1-r), where r=$<\mathbf{L.S}>/<n_h>$ and $n_h$ is the number of holes in the 5d manifold \cite{VDLXAS}. Using the XAS branching ratio of 2.6 and n$_h$=7 for Os$^{5+}$ gives $<\mathbf{L.S}>\approx 1$ in units of $\hbar^2$. This contrasts with $<\mathbf{L.S}>\approx 2$ for Ir$^{4+}$ in BaIrO$_3$ from XAS using the same method with BR=4 and n$_h$=5.[6] The reduced value of the expectation value of the SOC in NaOsO$_3$, compared to BaIrO$_3$, indicates SOC is not the explanation for the observed spin-phonon behavior. This suggests that an alternative phenomenon in 5d oxides, such as the extended electronic wavefunction, the route to achieve enhance spin-phonon coupling.

**Supplemental Figure S3**

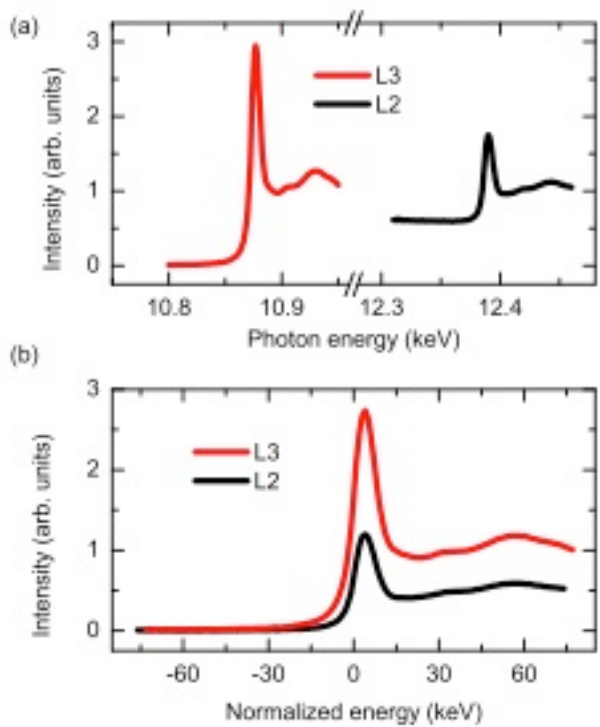

**S3 | X-ray near edge absorption spectroscopy (XANES) measurements investigating SOC. a** XANES results at the $L_2$ and $L_3$ absorption edges for $NaOsO_3$. **b** The XAS results shifted in energy to be centered at zero on their respective absorption energies and the intensities normalized to allow a comparison. The ratio of the integrated intensities is 2.6:1.